\documentclass[reprint,amsmath,amssymb,aps,longbibliography]{revtex4-1}
\usepackage{amsbsy}
\usepackage{amsmath}
\usepackage{amsfonts}
\usepackage{amsthm}
\usepackage[T1]{fontenc}
\usepackage{color}
\usepackage{textcomp}
\usepackage{float}
\usepackage{epstopdf}
\usepackage{textcomp}
\usepackage{hyperref}
\usepackage{dsfont}
\usepackage{enumitem}
\usepackage{etoolbox}
\usepackage{mathrsfs}
\usepackage{revsymb}
\usepackage{verbatim}

\usepackage{hyperref}

\usepackage{graphicx,caption,subcaption}
\usepackage{dcolumn}
\usepackage{bm}
\usepackage{comment}
\usepackage{physics}
\usepackage{tabu}
\usepackage{scrextend}
\usepackage{tabularx}
    \newcolumntype{L}{>{\raggedright\arraybackslash}X}

\theoremstyle{plain}

\theoremstyle{definition}
\newtheorem{definition}{Definition}[section]
\theoremstyle{remark}

\begin{document}

\title{Broadcasting of entanglement via orthogonal \& non-orthogonal state dependent cloners}

\author{Manish Kumar Shukla}
\affiliation{Center for Computational Natural Sciences and Bioinformatics, International Institute of Information Technology-Hyderabad, Gachibowli, Telangana-500032, India.} 
\author{Indranil Chakrabarty}
\affiliation{Center for Security, Theory and Algorithmic Research, International Institute of Information Technology-Hyderabad, Gachibowli, Telangana-500032, India.}
\author{Sourav Chatterjee}
\affiliation{Center for Computational Natural Sciences and Bioinformatics, International Institute of Information Technology-Hyderabad, Gachibowli, Telangana-500032, India.} 
\affiliation{SAOT, Erlangen Graduate School in Advanced Optical Technologies, Paul-Gordan-Strasse 6, 91052 Erlangen, Germany}
\affiliation{Raman Research Institute, Sadashivanagar, Bangalore, Karnataka-560080, India.}
\email[current affiliation \& email: ]{sourav.chatterjee@rri.res.in}


\begin{abstract}
In this work, we extensively study the problem of broadcasting of entanglement as state dependent versus state independent cloners. We start by re-conceptualizing the idea of state dependent quantum cloning machine (SD-QCM), and in that process, we introduce different types of SD-QCMs, namely, orthogonal and non-orthogonal cloners. We derive the conditions for which the fidelity of these cloners will become independent of the input state. We note that such a construction allows us to maximize the cloning fidelity at the cost of having partial information of the input state. 
In the discussion on broadcasting of entanglement, we start with a general two qubit state as our resource and later we consider a specific example of Bell diagonal state. We apply both state dependent and state independent cloners (orthogonal and non-orthogonal), locally and non locally, on input resource state and obtain a range for broadcasting of entanglement in terms of the input state parameters. Our results highlight several instances where the state dependent cloners outperform their state independent counterparts in broadcasting entanglement. Our study provides a comparative perspective on the broadcasting of entanglement via cloning in two qubit scenario, when we have some knowledge of the resource ensemble versus a situation when we have no such information. 
\end{abstract}
\maketitle

\section{Introduction}

\indent The principles \cite{heisenberg1927w} and resources \cite{einstein1935einstein, ollivier2001quantum, baumgratz2014quantifying} of quantum mechanics, on the one hand, gives us a significant advantage in accomplishing various information processing tasks \cite{bennett1984quantum, shor2000simple, ekert1991quantum, hillery1999m,sazim2015retrieving,adhikari2010probabilistic,ray2016sequential, bennett1993teleporting,horodecki1996teleportation, bennett1992communication, nepal2013maximally, horodecki2009quantum, bose1998multiparticle,zukowski1993event, sazim2013study, deutsch1996quantum} over their classical counterparts, while on the other hand, imposes strict restrictions on certain kind of operations which we can implement on the quantum states. In literature, these restrictions have been addressed in the form of various `No-go' theorems \cite{wootters1982single, pati2000impossibility, pati2006no,  chattopadhyay2006no, chakrabarty2007impossibility, zhou2006quantum, oszmaniec2016creating}.

Among these No-go theorems, the No-cloning theorem prohibits us from perfectly cloning an unknown quantum state \cite{wootters1982single}. However, it does not rule out the possibility to clone states approximately i.e. with a fidelity of less than unity \cite{wootters1982single, buvzek1996quantum, gisin1997optimal, buvzek1998universal, bruss1998optimal, cerf2000pauli, cerf2000nj, scarani2005quantum}. Approximate quantum cloning machines (QCMs) can further be classified into two types, namely (a) state dependent \cite{wootters1982single, buvzek1996quantum, bruss1998optimal, adhikari2006broadcasting} and (b) state independent \cite{buvzek1996quantum, buvzek1998universal} cloners. In our work, for the above classification, we have considered the set of symmetric cloners which produce both outputs with equal fidelity \cite{bruss1998optimal}. Recently, it was shown that a very high cloning fidelity can be obtained with such cloners via weak measurements \cite{wang2019high}. However, this is not the only set i.e. there exist asymmetric cloners having different fidelities for the two states at the output \cite{cerf2002security, ghiu2003asymmetric}. In state dependent quantum cloning machines (SD-QCMs), the performance of the cloning machine is dependent on the state to be cloned i.e. fidelities of cloned outputs are functions of the input state parameters, whereas for the state independent quantum cloning machines (SI-QCMs), the performance is independent of the input state parameters. The performance of SD-QCMs can be better than SI-QCMs for certain input states when a prior knowledge of the input ensemble is available, but on an average, SI-QCMs perform better than the state dependent cloners.

Entanglement \cite{einstein1935einstein}, which lies at the heart of quantum mechanics, acts as an invaluable resource in information processing \cite{gruska2003quantum}. In general, these resources are mixed entangled states \cite{werner1989quantum, horodecki2009quantum} in contrast to being pure entangled states \cite{white1999nonmaximally}. However, pure entangled resources achieve better efficiency in different information processing tasks. In literature, the process of distilling lesser number of pure entangled states from the available mixed entangled states has been studied quite extensively \cite{bennett1996concentrating, murao1998multiparticle}. This activity can be thought of as compression of entanglement. Interestingly, such a requirement is not unidirectional. In a network, there can be an exigency for producing more number of states (say with two qubits, can also work in higher dimensions) with lesser degree of entanglement, than having a single pair with higher degree of entanglement. The process of creating more number of lesser entangled pairs from an initial entangled pair can be viewed as decompression of entanglement. Such a decompression task can be achieved with more than one approach \cite{jain2017asymmetric}. However, when local and non-local cloning transformations are employed to achieve it, then in literature it is commonly referred to as `broadcasting of entanglement via cloning' \cite{buzek1997broadcasting, bandyopadhyay1999broadcasting, adhikari2006broadcasting}.

Bu{\v{z}}ek et. al. showed that though perfect broadcasting of entanglement is forbidden as a consequence of the no cloning theorem, partial decompression of initial quantum entanglement is possible, i.e., from a pair of entangled particles, two lesser entangled pairs can be obtained using cloning operations \cite{buzek1997broadcasting}. In this context, it was shown by Bandyopadhyay et. al. that only universal cloners (i.e. SI-QCMs), which have fidelity greater than $\frac{1}{2}(1+\sqrt{\frac{1}{3}})$, can broadcast entanglement \cite{bandyopadhyay1999broadcasting}. There also exists a bound on the number of copies that can be produced. This bound is two, if one uses local cloning operations and becomes six when one considers non-local cloning operations \cite{bandyopadhyay1999broadcasting}. Recently, it was shown that it is impossible to even partially broadcast quantum correlation that goes beyond entanglement (namely, discord \cite{henderson2001classical, ollivier2001quantum}), by using either local or non-local, symmetric \cite{chatterjee2016broadcasting} or asymmetric \cite{jain2017asymmetric} cloners. 

In this work, in section (II) we first introduce and discuss various types of cloning, namely, state dependent and state independent versions of local and non-local operations. We introduce two methods of constructing cloning transformations, namely, orthogonal and non-orthogonal cloning transformations. We also compare the performances of the above stated cloning operations in terms of distortion produced while cloning. In section (III), we develop the utility of cloning transformations into broadcasting of entanglement via both local and non-local cloning approaches. Lastly, in the final subsection, we study the problem of broadcasting of entanglement with an example of the Bell-diagonal state (BDS) \cite{bell-diagonal2010}.

\section{Orthogonal and Non-Orthogonal Cloning}

Perfect cloning is not possible according to the No-cloning theorem \cite{wootters1982single}. However, it never rules out the possibility to clone quantum states approximately with a fidelity $F$ less than unity\cite{buvzek1996quantum}, given by,
\begin{equation}
F = \left\langle \Psi|\rho^{out}|\Psi\right\rangle, 
\end{equation}
where $\left|\Psi\right\rangle$ refers to the state to be cloned at the input port of the cloner and $\rho^{out}$ is the state obtained at its output port after applying the cloning transformation. 

\subsection{Orthogonal cloning}

Among all the symmetric cloning machines available in literature for qubits, the  B-H cloning machine ($U_{bh}$) is optimal \cite{gisin1997optimal}. It is an $M$-dimensional quantum copying transformation acting on a state $\left|\Psi_i\right\rangle_{a_0}$ (where $i$ $\in$ \{1, ..., $M$\}). This state is to be copied on a blank state $\left|\Sigma\right\rangle_{a_1}$. Initially, the copying machine was prepared in state $\left|X\right\rangle_x$, which after being applied on the inputs gets transformed into another set of state vectors $\left|X_{ii}\right\rangle_x$ and $\left|Y_{ij}\right\rangle_x$ (where $i,\,j$ $\in$ \{1, ..., $M$\}). Here, $\ket{\Psi} = \sum_{i=1}^{M}\alpha_i\ket{\Psi_i}$, where $\ket{\Psi_i}$ are the basis vectors of the m qubit system with dimensions $M = 2^m$ and $\alpha_i$ represents the probability amplitude, hence $\sum_{i=0}^{M} \alpha^{2}_i = 1$. Here, the modes $a_0$, $a_1$ and $x$ represent the input, blank and machine qubits, respectively. In this case, these transformed machine state vectors $\left(\left|X_{ii}\right\rangle,\,  \left|Y_{ij}\right\rangle\right)$ are elements of the orthonormal basis set in the $M$-dimensional space. The transformation scheme $U_{bh}$ is given by \cite{buvzek1998universal},
\begin{eqnarray}
&& U_{bh}\left|\Psi_i\right\rangle_{a_0} \left|\Sigma\right\rangle_{a_1} \left|X\right\rangle_x \rightarrow  c\left|\Psi_i\right\rangle_{a} \left|\Psi_i\right\rangle_{b} \left|X_{ii}\right\rangle_x \nonumber\\
&& +d\displaystyle \sum_{j\neq i}^{M} \left(\left|\Psi_i\right\rangle_{a} \left|\Psi_j\right\rangle_{b} +\left|\Psi_j\right\rangle_{a}\left|\Psi_i\right\rangle_{b}\right) \left|Y_{ij}\right\rangle_x,
\label{eq:B-H_gen_transform}
\end{eqnarray}
where $i,\:j$ $\in$ $\{1,...,M\}$. The coefficients $c$ and $d$ are the probability amplitudes which can take real values. These values are associated with the probability of success and failure of redistributing information. We call this kind of cloning machine, where the the machine states are orthonormal, as Orthogonal Cloning Machine (OCM). We can find the relation between coefficients $c$ and $d$ by the unitarity condition \cite{buvzek1999universal},
\begin{equation}
c^2 = 1 - 2(M-1)d^2.
\label{eq:relationBetweenCAndD}
\end{equation}
So, we have only one independent machine state parameter, either $c$ or $d$, which can take any value between $0$ to $1$. The complete output state obtained after tracing out the machine state is given by,
\begin{eqnarray}
&&\rho_{ab}^{out} = \sum_{i=1}^{M} (c^2 \alpha_i\alpha_i^{*}(\ket{\Psi_i}_a\bra{\Psi_i}\otimes \ket{\Psi_i}_b\bra{\Psi_i})\nonumber\\ 
&&+ c d  \sum_{j=1, j\ne i}^{M} (\alpha_i \alpha_j^{*}  \ket{\Psi_i}\ket{\Psi_i} \bra{\Phi_{ij}} + \alpha_i^{*} \alpha_j  \ket{\Phi_{ij}}  \bra{\Psi_i}\bra{\Psi_i})\nonumber\\
&&+d^2 \sum_{j=1}^{M}(\alpha_i \alpha_j^{*}\sum_{k=1,k\ne i, k\ne j}^{M} \ket{\Phi_{jk}}\bra{\Phi_{ik}})).
\label{eq:mdimensioncompleteOrthogonal}
\end{eqnarray}
Here, $\ket{\Phi_{in}}=(\ket{\Psi_i}_a\ket{\Psi_n}_b+\ket{\Psi_n}_a\ket{\Psi_i}_b)$, where $n \in \{j,k\}$. The individual cloned states are given by,
\begin{eqnarray}
&& \rho_a^{out} = \rho_b^{out} =  \sum\limits_{i=1}^M (c^2 \alpha_i \alpha_i^{*} \ket{\Psi_i}\bra{\Psi_i} \nonumber\\
&& + c d \sum\limits_{j=1,i \ne j }^M  (\alpha_i \alpha_j^{*}\ket{\Psi_i}\bra{\Psi_j}+\alpha_i^{*} \alpha_j\ket{\Psi_j}\bra{\Psi_i})\nonumber\\
&& +  d^2 \sum\limits_{j=1,j \ne i}^M \alpha_i \alpha_i^{*}(\ket{\Psi_i}\bra{\Psi_i} +  \ket{\Psi_j}\bra{\Psi_j}).
\label{eq:mdimensiontracedOrthogonal}
\end{eqnarray}

\subsection{Non-orthogonal cloning}
Another way to define the general M dimensional quantum cloning machine, from which we can easily derive the state dependent cloner, is by substituting $c=d=1$ in Eq.~\ref{eq:B-H_gen_transform} . The cloning transformation is then given by,
\begin{equation}
\begin{split}
U_{bh}\ket{\Psi_i}_{a_0}\ket{\Sigma}_{a_1}\ket{X}_{x} \rightarrow \ket{\Psi_i}_{a}\ket{\Psi_i}_{b}\ket{X_{ii}}_{x}\\ + \sum_{j=1,j\ne i}^{M}(\ket{\Psi_i}_{a}\ket{\Psi_j}_{b} + \ket{\Psi_j}_{a}\ket{\Psi_i}_{b})\ket{Y_{ij}}_{x}.
\end{split}
\label{eq:cloner}
\end{equation}
We have introduced the non-orthogonality such that the following unitarity conditions are obeyed,
\begin{equation}
\begin{split}
\braket{X_{ii}}{X_{ii}} = 1-2(M-1)\lambda, \\
\braket{X_{ii}}{X_{jj}} = 0; \text{with $i$} \ne j, \\
\braket{Y_{ij}}{Y_{ij}} = \lambda, \\
\braket{X_{ii}}{Y_{ij}} = 0, \\
\braket{Y_{ij}}{Y_{kl}} = 0; \text{with $i$} \ne k, \\
\braket{X_{ii}}{Y_{jk}} = \mu/2; \text{with $i$} \ne j.
\end{split}
\label{eq:unitarity}
\end{equation}
Henceforth, we call this kind of machine as Non Orthogonal Cloning Machine (NOCM). In this case, both these quantities $\lambda$ and $\mu$ are independent and hence we have two machine parameters. However, Schwartz inequality 
imposes restrictions on the physically permitted values of $\lambda$ and $\mu$,

\begin{equation}
\begin{split}
|\braket{X_{ii}}{Y_{jk}}|^2 \leq \braket{X_{ii}}{X_{ii}} \braket{Y_{jk}}{Y_{jk}}\\
\implies |\frac{\mu}{2}|^2 \leq \lambda(1-2(M-1)\lambda).
\end{split}
\label{eq:schwartzNonOrthogonal}
\end{equation}
From Eq.~\ref{eq:unitarity} we can say that $\lambda$ can take values $0 \leq \lambda \leq \frac{1}{2(M-1)}$ and accordingly Eq.~\ref{eq:schwartzNonOrthogonal} will restrict the values of $\mu$.

The complete density matrix of the combined output state after tracing out machine states is given by,
\begin{eqnarray}
&& \rho_{ab}^{out} = (1-2(M-1)\lambda)\sum_{i=1}^{M}\alpha_i\alpha_i^{*}(\ket{\Psi_i}_a\bra{\Psi_i}\otimes \ket{\Psi_i}_b\bra{\Psi_i})\nonumber\\
&& + \lambda \sum_{i=1}^{M}\alpha_i\alpha_i^{*}  \sum_{j=1, j\ne i}^{M}\ketbra{\chi_{ij}}{\chi_{ij}}
+ \frac{\mu}{2} \sum_{i=1}^{M}\alpha_i \sum_{j=1, j\ne i}^{M}\alpha_j^{*}\nonumber\\
&& \sum_{k=1,k\ne i}^{M}(\ket{\Psi_i}_a\ket{\Psi_i}_b\bra{\Phi_{jk}} + \ket{\Phi_{jk}}\bra{\Psi_i}_a\bra{\Psi_i}_b).
\label{eq:mdimensioncompleteNonOrthogonal}
\end{eqnarray}
Here, $\ket{\chi_{ij}}=(\ket{\Psi_i}_a\ket{\Psi_j}_b+\ket{\Psi_j}_a\ket{\Psi_i}_b)$ and $\ket{\Phi_{jk}}=(\ket{\Psi_j}_a\ket{\Psi_k}_b+\ket{\Psi_k}_a\ket{\Psi_j}_b)$. After tracing out one of the subsystem, we get the new cloned state at the output port  as,
\begin{eqnarray}
&& \rho_a^{out} = \rho_b^{out} =  (1-2(M-1)\lambda) \sum\limits_{i=1}^M \alpha_i \alpha_i^{*} \ket{\Psi_i}\bra{\Psi_i} \nonumber\\
&& +\frac{\mu}{2} \sum\limits_{i=1}^M \alpha_i \sum\limits_{j=1,i \ne j }^M \alpha_j^{*} (\ket{\Psi_i}\bra{\Psi_j}+\ket{\Psi_j}\bra{\Psi_i})\nonumber\\
&& + \lambda  \sum\limits_{i=1}^M \alpha_i \alpha_i^{*} \sum\limits_{j=1,j \ne i}^M (\ket{\Psi_i}\bra{\Psi_i} +  \ket{\Psi_j}\bra{\Psi_j}).
\label{eq:mdimensiontracedNonOrthogonal}
\end{eqnarray}

To calculate the optimal value of the machine parameter $\lambda$, we make $\frac{\partial D_{ab}}{\partial \lambda} = 0$ and obtain the $\lambda$ for which the value of $D_{ab}$ (distortion due to cloning) is minimum. This ensures that the machine parameters selected causes minimal distortion to the ideal output. Now, when we have a cloning transformation for $m$ qubits, we can apply these to study the effect of cloning on teleportation, broadcasting, discord \cite{chatterjee2016broadcasting} and coherence \cite{sharma2017broadcasting}. 

\subsection{State dependent (SD) cloner}
In general, we can copy an unknown quantum state with some imperfection which can be associated with the distortion on the ideal output state. We introduce two quantities, $D_{ab} =  Tr[\rho_{ab}^{(out)} -\rho_{a}^{(id)}\otimes\rho_{b}^{(id)} ]^2$ and  $D_{a} =  Tr[\rho_{a}^{(out)} -\rho_{a}^{(id)}]^2$ to quantify the amount of distortion in the combined system and the individual system as a result of cloning, respectively. Here, $\rho_a^{(id)}$ and $\rho_b^{(id)}$ represent the outputs for ideal cloning operation, that is, $\rho_a^{(id)}$ = $\rho_b^{(id)}$ = $\rho_a^{(in)}$. We use these distance based measures to quantify the performance of cloning machines. We can classify cloning machines into two groups based on the kind of imperfection (distortion produced) they create. If the
performance of the cloning machine is dependent on the input state i.e., if the cloner
performs good for some  states  and  bad  for  some  other  states,  the  cloning  machine  is  called  state  dependent cloner (SD-QCM). In general, if we select any value for the machine parameters ($d$ in case of OCM and $\lambda$, $\mu$ for NOCM), the performance of the machine becomes dependent on the state being cloned. However, there are specific values of machine parameters which make the performance independent of the input state. We describe this in the following subsection.

\subsection{State independent (SI) cloner}
If a cloning machine performs the same for any input state it is called state independent (universal) quantum cloning machine (SI-QCM). 

In case of NOCM, to get the relation between $\lambda$ and $\mu$, we use the following condition $\frac{\partial D_{a}}{\partial \lvert \alpha_i\rvert^2} = 0 $, which basically sets $D_{a}$ independent of the input state parameters. For the M dimensional case, we get 
\begin{equation}
\mu = 1 - M\lambda.
\label{eq:relationLambdaMew}
\end{equation}
When we substitute $\mu$ with $1-M\lambda$, we find that the distortion $D_a$ reduces to 
\begin{equation}
D_a = M(M-1)\lambda^2.
\label{eq:distortionA}
\end{equation}
It is evident that the performance is independent of input parameter $\alpha_i$ and it is only dependent on $\lambda$. However, the value of $\lambda$ is constrained by Eq.~\ref{eq:schwartzNonOrthogonal}. If we substitute $\mu=1-M\lambda$ in Eq.~\ref{eq:schwartzNonOrthogonal},
\begin{equation}
\bigg( \frac{1-M\lambda}{2}\bigg)^2 \leq \lambda(1-2(M-1)\lambda).
\label{eq:limitondimentinality}
\end{equation}
We find that $M\leq3$, for $\lambda$ to have any real solution. This implies that we cannot have a NOCM which is state independent beyond three dimensions, i.e. it is only restricted to a single qubit and qutrit scenario. 

In case of OCM, when we apply the condition of state independence, $\frac{\partial D_{a}}{\partial \lvert \alpha_i\rvert^2} = 0 $, we get the values of coefficients $c$ and $d$ as follows,
\begin{equation}
d^2 = \frac{1}{2(M+1)}. 
\label{eq:stateindependence}
\end{equation}
Thus for $OCM$ it reduces to the M-dimensional B-H QCM\cite{buvzek1999universal}.

\begin{figure}[h]
\begin{center}
\includegraphics[scale=0.72]{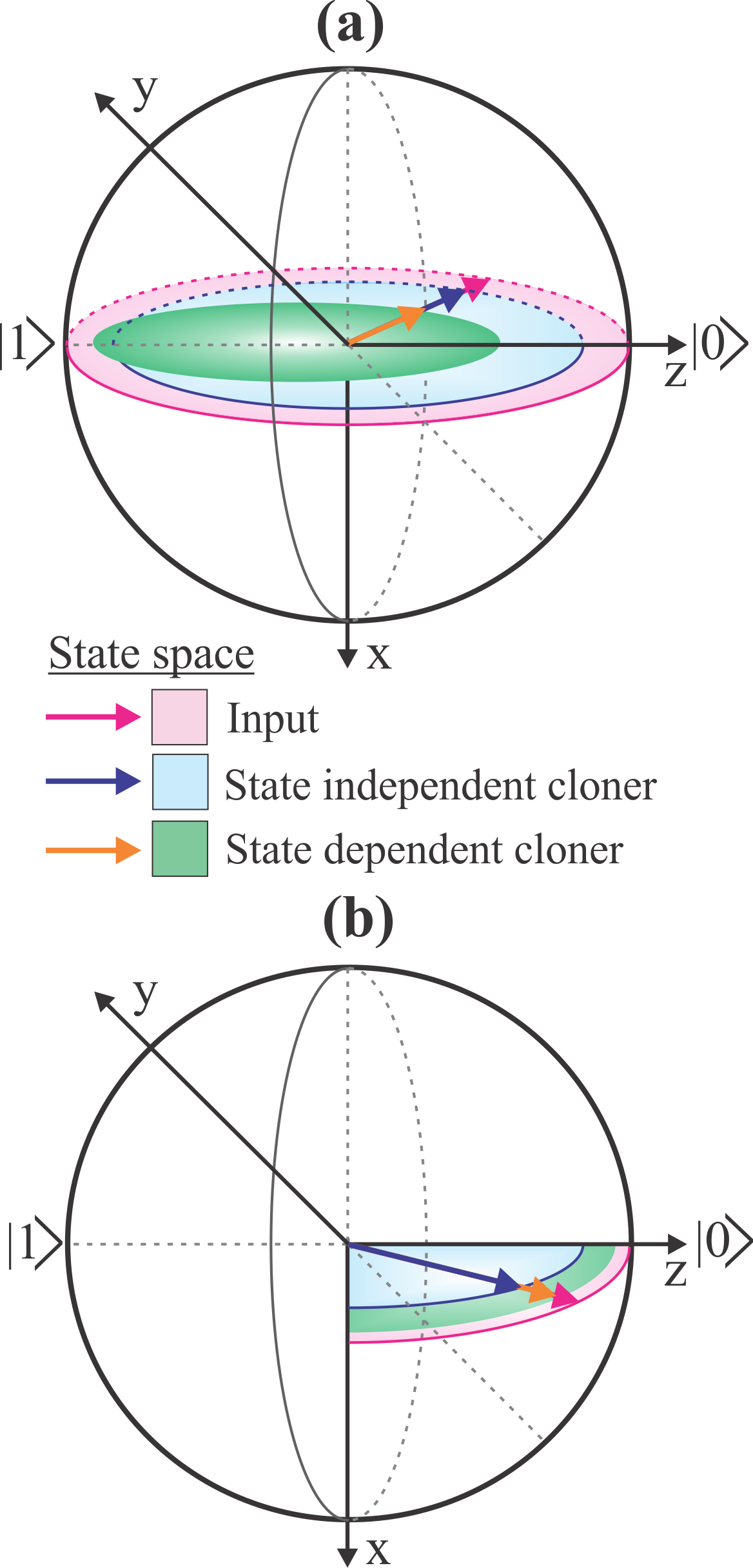}
\end{center}
\caption{\noindent \scriptsize
A schematic comparing the performance of a state dependent (SD) versus a state independent (SI) cloner on an input qubit, inside the Bloch sphere framework. The x, y and z basis vectors are essentially the Pauli matrices. In sub-figure (a), the pink circle traced azimuthally around the x-basis represents the input state space of the qubit which is to be cloned. As inputs, we restrict our illustration to pure states so the pink vector can only move along its circumference (i.e. solid pink boundary). The output state space of the cloned qubit is depicted with a blue circle for the SI cloner and with a green elliptical envelope for the SD cloner. Similarly, in sub-figure (b), a corresponding comparison is sketched considering only one quadrant of the previous input state space. It symbolizes the case when prior information on a restricted range of input state parameters is available during the preparation of the SD cloner.  \label{fig:cloneschematic}}
\end{figure}

In Fig.~\ref{fig:cloneschematic} we illustrate the performance of a state dependent (SD) cloner against a state independent (SI) cloner, for a given class of pure input states, through a schematic representation on the Bloch sphere. For a comparative analysis, we present two cases in the two sub-figures: (a) when the class of input states is given (i.e. $\left|in\right\rangle = \alpha \left|0\right\rangle + \sqrt{1-\lvert\alpha\rvert^2} \left|1\right\rangle$) but there exists no prior information on the range of the probability distribution from which the inputs are being chosen (i.e. $0\leq \lvert\alpha\rvert^2 \leq 1$); and (b) when the class of input states is provided (i.e. $\left|in\right\rangle = \alpha \left|0\right\rangle + \sqrt{1-\lvert\alpha\rvert^2} \left|1\right\rangle$) as well as more knowledge about the restricted range of the probability distribution from which the inputs are being chosen is available (i.e. $\xi\leq \lvert\alpha\rvert^2 \leq \tau$ where $\xi>>0$  and $\tau<<1$) during the preparation of the SD cloner. The former and latter input cases are sketched with a complete pink circular region about the x-basis in Fig.~\ref{fig:cloneschematic}(a) and with a pink quadrant in Fig.~\ref{fig:cloneschematic}(b) respectively. In both sub-figures, the dark pink input state vector can only move along the circumference with a fixed radius as we restrict our illustration to only pure input states. The SI cloner clones all inputs states with an optimal cloning fidelity of $\frac{5}{6}$ \cite{gisin1997optimal}. So in sub-figure (a), corresponding to the pink input state space, the cloned output state space for the SI cloner is depicted by the (larger) blue circular region having a radius (shown with a dark blue vector) equivalent to $\frac{5}{6}^{\text{th}}$ of the (dark pink) input state vector. However for the SD cloner, the cloned output state space is given by the (smaller) green elliptical regime and is traced out by an orange vector of variable length. In principle the fidelity of cloning will always be less than unity, so the outputs would become mixed states. Hence, the output vectors (dark blue and orange) are shown to shrink along the direction of the dark pink input state vector according to their respective cloning fidelities \cite{bruss1998optimal}. In sub-figure (a), their is no prior information about the input state parameters, and hence the overall area of the blue region is larger than the green region, showing that the average performance of SI cloners is better. In sub-figure (b), all other descriptions of the inputs and the cloners remain consistent with those of sub-figure (a); except that here we constrict our illustration to only a quadrant, symbolizing that now we have prior information about the restricted range of the possible input states. In this case, we note that the green elliptical region has a larger area on average than the blue one, claiming that the SD cloner outperforms the SI cloner under such circumstances.

\subsection{Cloning of a general two qubit state}
 A general two qubit state $\rho_{12}$, represented in terms of Bloch vectors and correlation matrix can be expressed as, 
\begin{equation}
\begin{split}
\rho_{12} = \frac{1}{4}\Big[\mathds{I}_4 + \sum_{u=1}^3(x_u\sigma_u\otimes\mathds{I}_2 + y_u\mathds{I}_2\otimes\sigma_u) \\
+ \sum_{u,v=1}^3t_{uv}\sigma_u\otimes\sigma_v \Big] = \big\{\vec{x},\vec{y},T\big\}, \label{eq:rowgen}
\end{split}
\end{equation}
where $x_u = Tr[\rho_{12}(\sigma_u\otimes\mathds{I}_2)]$, $y_u = Tr[\rho_{12}(\mathds{I}_2\otimes\sigma_u)]$, $t_{uv} = Tr[\rho_{12}\sigma_u\otimes\sigma_v]$, and $\sigma_u$ are Pauli matrices. $\mathds{I}_n$ is the identity matrix of order $n$ and ($u$, $v$) $\in$ $\{1,2,3\}$. In the simplified expression, $\vec{x}$, $\vec{y}$ are Bloch vectors and $T=[t_{uv}]$ is the correlation matrix. 

\subsubsection{Local cloner}
Let us consider a situation where qubit $1$ is with Alice and qubit $2$ is with Bob. We apply two local cloners (as given by Eq.~\ref{eq:B-H_gen_transform} with $M=2$), $U_1 \otimes U_2$, on $\rho_{12}$ (as defined in Eq.~\ref{eq:rowgen}) with blank state $\ket{\Sigma}_3$ on Alice's side and $\ket{\Sigma}_4$ on Bob's side. So the two qubit blank state can then be jointly expressed as $\mathscr{B}_{34} = \ketbra{\Sigma}{\Sigma} \otimes \ketbra{\Sigma}{\Sigma}$. On tracing out the non-local qubits in each case and the machine states $\mathscr{M}_{56} = \ketbra{X}{X} \otimes \ketbra{X}{X}$, we get the local output states: 
\begin{eqnarray}
&& \rho_{13}^{out} = Tr_{2456}[U_1\otimes U_2(\rho_{12}\otimes\mathscr{B}_{34}\otimes\mathscr{M}_{56})U_2^\dagger\otimes U_1^\dagger]\, \text{and} \nonumber\\
&& \rho_{24}^{out} = Tr_{1356}[U_1\otimes U_2(\rho_{12}\otimes\mathscr{B}_{34}\otimes\mathscr{M}_{56})U_2^\dagger\otimes U_1^\dagger].
\label{eq:localoutputs}
\end{eqnarray}
In a similar approach, the non-local output states are given by, 
\begin{eqnarray}
&& \rho_{14}^{out} = Tr_{2356}[U_1\otimes U_2(\rho_{12}\otimes\mathscr{B}_{34}\otimes\mathscr{M}_{56})U_2^\dagger\otimes U_1^\dagger]\, \text{and} \nonumber\\
&& \rho_{23}^{out} = Tr_{1456}[U_1\otimes U_2(\rho_{12}\otimes\mathscr{B}_{34}\otimes\mathscr{M}_{56})U_2^\dagger\otimes U_1^\dagger]. 
\label{eq:nonlocalouputs}
\end{eqnarray}

If we use non-orthogonal cloning transformation as given in Eq.~\ref{eq:cloner}, the local and the non-local outputs obtained after cloning as produced using Eqs. (\ref{eq:localoutputs}, \ref{eq:nonlocalouputs}) are given by,
\begin{equation}
\begin{split}
\rho_{13}^{out} = \bigg\{\mu\vec{x},\mu\vec{x},T_{l,l}^{sd}\bigg\}, \rho_{24}^{out} = \bigg\{\mu\vec{y},\mu\vec{y},T_{l,l}^{sd}\bigg\},
\end{split}
\end{equation}
and
\begin{equation}
\rho_{14}^{out} = \rho_{23}^{out} = \bigg\{\mu\vec{x},\mu\vec{y},\mu^2 T\bigg\},
\label{eq:localnonorthooutput}
\end{equation}
respectively. Here $T_{l,l}^{sd}$ = diag[$2\lambda$ ,$2\lambda$, 1-$4\lambda$]. Recall that $\mu = 1-M\lambda$ was the machine parameter for our universal cloner (SI-QCM).
In case of orthogonal cloning transformation, as given in Eq.~\ref{eq:B-H_gen_transform}, the local and the non-local outputs obtained after cloning are given by,
\begin{equation}
\begin{split}
\rho_{13}^{out} = \bigg\{\chi_{l}\vec{x},\chi_{l} \vec{x},T_{l,l}^{osd}\bigg\}, \rho_{24}^{out} = \bigg\{\chi_{l}\vec{y},\chi_{l} \vec{y},T_{l,l}^{osd}\bigg\},
\end{split}
\end{equation}
\begin{equation}
\rho_{14}^{out} = \rho_{23}^{out} = \bigg\{\chi_{l}\vec{x},\chi_{l} \vec{y}, T_{nl,l}^{osd}\bigg\},
\label{eq:localorthooutput}
\end{equation}

where $T_{l,l}^{osd}$ = $\text{diag}[2d^2, 2d^2, 1-4d^2]$, $\chi_{l} = diag[2cd, 2cd, 1-4d^2]$ and  \[
   T_{nl,l}^{osd}=
  \left[ {\begin{array}{ccc}
   4c^2d^2T_{11} & 4c^2d^2T_{12} & 2c^{\frac{3}{2}}dT_{13}\\
   4c^2d^2T_{21} & 4c^2d^2T_{22} & 2c^{\frac{3}{2}}dT_{23}\\
   2c^{\frac{3}{2}}dT_{31} & 2c^{\frac{3}{2}}d T_{32} & c^4T_{33}\\
  \end{array} } \right].
\]

\subsubsection{Non-local cloner}
Next, we consider the case when the two qubit state $\rho_{12}$  (as defined in Eq.~\ref{eq:rowgen}) is being cloned with higher dimensional non-local  cloners. In case of non-local cloning, we apply a higher dimensional unitary transformation $U_{12}$ (as given by Eq.~\ref{eq:B-H_gen_transform} with $M=4$), on the combined two qubit state $\rho_{12}$ instead of two separate local cloners. The output states obtained in this process are then given by, 
\begin{eqnarray}
\rho _{ij}^{out} = Tr_{kl56}[U_{12}\left(\rho_{12}\otimes\mathscr{B}_{34}\otimes\mathscr{M}_{56}\right)U^\dagger_{12}],\\ 
\forall\, i\neq j \neq k \neq l\,\, \&\,\, i,\, j,\, k,\, l\, \in\, \left\{1,2,3,4\right\};\nonumber
\end{eqnarray}
where $\mathscr{B}_{34}$ is the two qubit blank state and $\mathscr{M}_{56}$ represent the two qubit machine state. In case of the non-orthogonal cloning, the local and the non-local output states are given by, 

\begin{equation}
\begin{split}
\rho_{13}^{out} = \bigg\{\mu\vec{x},\mu\vec{x},T_{l,nl}^{sd}\bigg\}, \rho_{24}^{out} = \bigg\{\mu\vec{y},\mu\vec{y},T_{l,nl}^{sd}\bigg\},
\end{split}
\end{equation}
\begin{equation}
\rho_{12}^{out} = \rho_{34}^{out} = \bigg\{\mu\vec{x},\mu\vec{y},\mu T\bigg\}.
\label{eq:nonlocalnonorthooutput}
\end{equation}
Here, the matrix $T_{l,nl}^{sd}$ is a $3\times3$ diagonal matrix, with the diagonal elements being $2\lambda, 2\lambda$ and $1-8\lambda$. Here, $T$ is the same input state correlation matrix. 

In case of orthogonal cloning transformation, the local and the non-local outputs obtained after cloning are given by,
\begin{equation}
\begin{split}
\rho_{13}^{out} = \bigg\{\chi_{nl}\vec{x},\chi_{nl} \vec{x},T_{l,nl}^{osd}\bigg\}, \rho_{24}^{out} = \bigg\{\chi_{l,nl}\vec{y},\chi_{nl} \vec{y},T_{l,nl}^{osd}\bigg\},
\end{split}
\end{equation}
\begin{equation}
\rho_{12}^{out} = \rho_{34}^{out} = \bigg\{\chi_{nl}\vec{x},\chi_{nl} \vec{y}, 2d(c+d) T_{nl,nl}^{osd}\bigg\},
\label{eq:nonlocalorthooutput}
\end{equation}

where $T_{l,nl}^{osd}$ = $\text{diag}[2d^2,\,2d^2,\,1-8d^2]$, $\chi_{nl} = \text{diag}[2d(c+d)d, 2d(c+d), 1-4d^2]$ and  $T_{nl,nl}^{osd}$ is same as the input state correlation matrix $T$, except the last entry, $T'_{33} = \frac{(1 - 4 d^2)}{2d(c+d)})T_{33}$.

\subsection{Comparative analysis of orthogonal versus non-orthogonal cloner}
If we look at the average performance on the entire range for the allowed input state parameters, the state independent (universal) cloner performs the best. Incidentally, the state independent (SI) orthogonal and non-orthogonal cloners perform the same and the values of their machine parameters become $\lambda = \frac{1}{6}, \mu = \frac{2}{3}$ and $d = \sqrt{\frac{1}{6}}$. However, when we have prior partial information about the state, the state dependent (SD) cloner seems to perform better. 

To demonstrate this, we take an example of a single qubit pure state $\ket{\Omega} = cos[\frac{\theta}{2}] \ket{0} + e^{i\phi} sin[\frac{\theta}{2}]\ket{1}$. If we do not have any information about the state, then it would mean $\theta$ can take any value between $0$ to $\pi$ and $\phi$ can take values between $0$ to $2\pi$. One of the ways in which we can have partial information is by restricting the state parameter $\theta$ to lie between a sub range of $0$ to $\pi$. One way to do this is to say that $\theta$ lies in the range $z$ to $\pi - z$, where z can take any value between $0$ to $\frac{\pi}{2}$. Higher value of $z$ represents lesser uncertainty in system and consequently more knowledge (or prior information) about the input state. 

\begin{figure}[h]
\begin{center}
\includegraphics[scale=0.62]{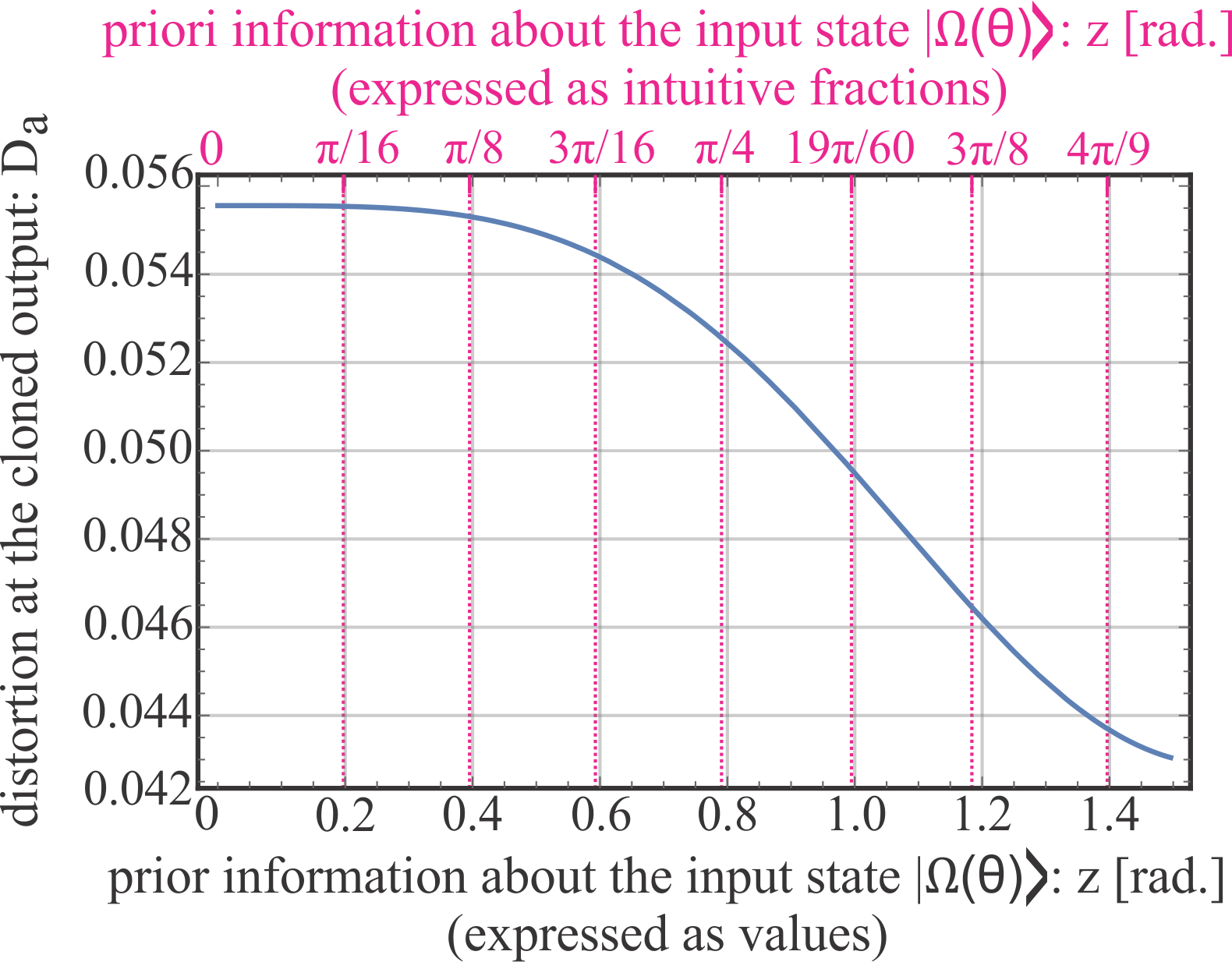}
\end{center}
\caption{\noindent \scriptsize
Plot comparing the amount of distortion ($\text{D}_{\text{a}}$) produced in the imperfect version of locally cloned output from the ideal output (or the input state) against the prior information (expressed in $z$ radians) available about the input state $\ket{\Omega} = cos[\frac{\theta}{2}] \ket{0} + e^{i\phi} sin[\frac{\theta}{2}]\ket{1}$, where $z \leq \theta \leq \pi-z$.}
\label{fig:localwithinfo}
\end{figure}

We find the optimal fidelity in case of the SD (orthogonal and non-orthogonal) cloning by varying the value of $z$. Interestingly, the performance of an optimal version of orthogonal and non-orthogonal SD cloner, on a single input qubit (i.e. when operated locally), is the same. We also note that the distortion ($\text{D}_{\text{a}}$) decreases when we increase the value of $z$, which implies that more the prior information about the state, lesser is the distortion in the cloned output from the ideal one. This relationship for the SD local cloner can be observed from the plot in Fig.~\ref{fig:localwithinfo}.
An important thing to note here is that even if we know the exact value of $\theta$, i.e. $z = \frac{\pi}{2}$, $\text{D}_{\text{a}}$ does not vanish to $0$, but takes a finite value. This is because the value of $\phi$ remains still unknown.

\begin{figure}[h]
\begin{center}
\includegraphics[scale=0.63]{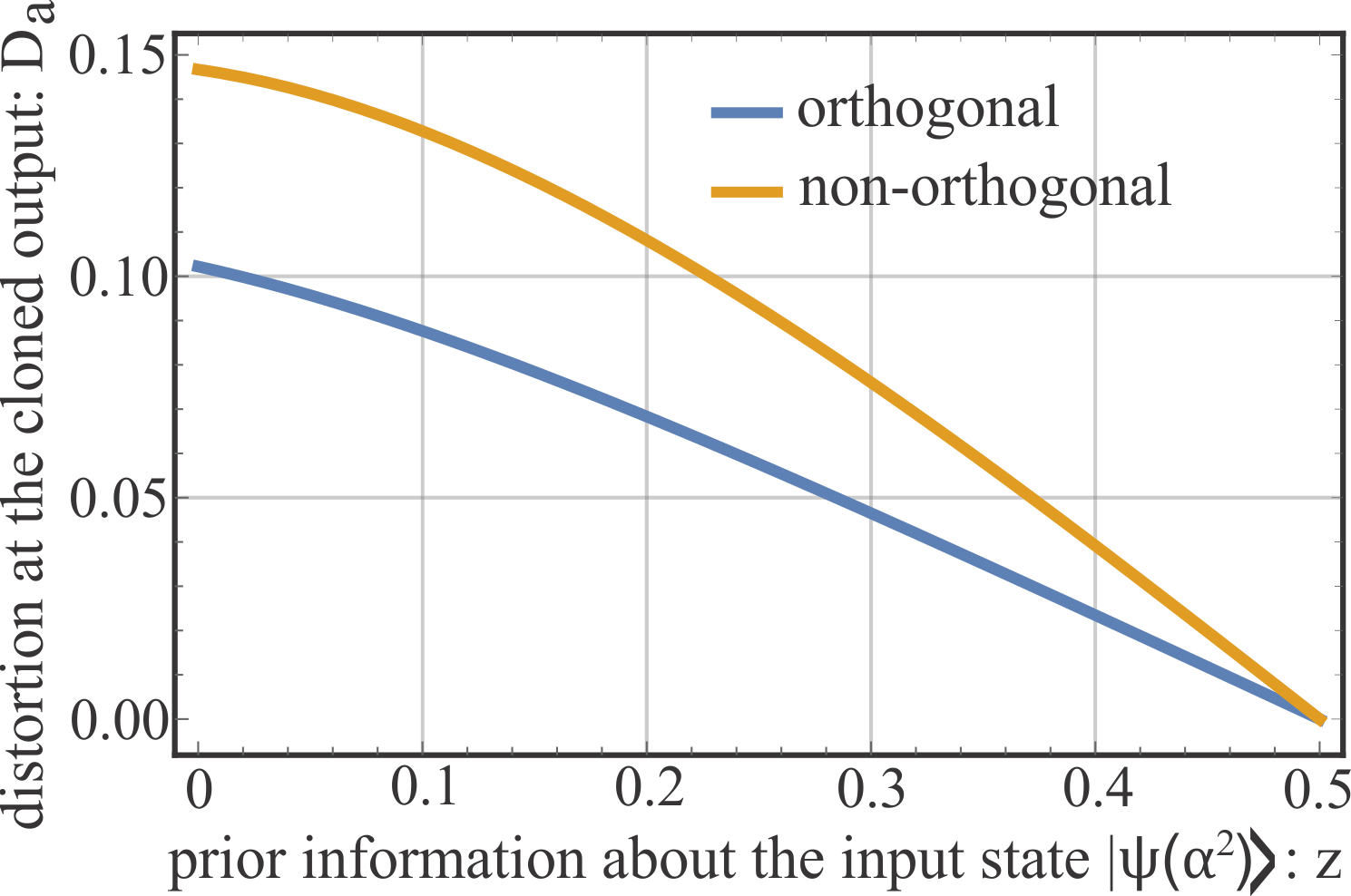}
\end{center}
\caption{\noindent \scriptsize
Plot comparing the amount of distortion ($\text{D}_{\text{a}}$) produced in the imperfect version of non-locally cloned output from the ideal output (or the input state) against the prior information (expressed in $z$) available about the input state $\ket{\Psi} = \alpha \ket{00} + \sqrt{1-\alpha^2}\ket{11}$, where $\alpha^2$ takes values between [$z$, $1-z$]. We observe that the orthogonal cloner performs better than its non-orthogonal version.}
\label{fig:nonlocalwithinfo}
\end{figure}

The non-local case has further interesting results. This is because of the stricter restrictions set on machine parameters in the case of non-orthogonal cloning. As a consequence of Eq.~\ref{eq:limitondimentinality}, the universal (or SI) version does not exist for non-local non-orthogonal cloners. Consequently, for such non-local class of non-orthogonal SD cloners the worst performance is produced when no prior information about the input state is available. However in case of the non-local orthogonal cloners, the universal (or SI) version does exist and the optimal distortion achieved with it is $0.12$. The relationship between non-local SD orthogonal and non-orthogonal cloners is plotted in Fig.~\ref{fig:nonlocalwithinfo}. It is clear from the figure that the SD orthogonal cloner will perform much better than its SI counterpart as more and more prior information is made available. Interestingly, here we also find that the non-local SD orthogonal cloners always perform better than the non-orthogonal ones, irrespective of how much prior information about the system has been provided. This is because of the additional restrictions imposed by Schwartz inequality (as per Eq.~\ref{eq:schwartzNonOrthogonal}) on the allowed values of $\lambda$ and $\mu$ in case of NOCM. 

\section{Broadcasting of entanglement via cloning}
\subsection{Broadcasting}
Broadcasting of entanglement refers to a situation where, from a given entangled state, we can create lesser entangled states by applying unitary transformations, both locally and non-locally. There are several ways to implement this however, the most popular way is the one where we use both local and non-local cloning machines.

Our basic aim is to broadcast the maximal amount of entanglement present in the given input state to many pairs. The procedure we follow to achieve this is to take an entangled state $\rho_{12}$ (shared between two parties say Alice and Bob) along with a two qubit blank state, apply the cloning transformation and finally trace out the machine states to obtain the combined output state $\rho_{1234}$. 
\theoremstyle{definition}
\begin{definition}{\textit{Optimal broadcasting:}}
An entangled state $\rho_{12}$ is said to be optimally broadcast after the application of cloning operation $U_{cl}$ ($U_1 \otimes U_2$ in case of local cloning and $U_{12}$ in case of non-local cloning) given by Eq.~\ref{eq:cloner}, if for some values of the input state parameters,
\begin{itemize}
\item the non-local output states between Alice and Bob:
\begin{equation}
\begin{split}
\rho_{14}^{out} = Tr_{23}[U_{cl}\rho_{12}],
\rho_{23}^{out} = Tr_{14}[U_{cl} \rho_{12}];\, \text{or} \\
\rho_{12}^{out} = Tr_{34}[U_{cl}\rho_{12}],
\rho_{34}^{out} = Tr_{12}[U_{cl} \rho_{12}] 
\end{split}
\end{equation}
 are inseparable, when
\item the local output states of Alice and Bob:
\begin{equation}
\begin{split}
\rho_{13}^{out} = Tr_{24}[U_{cl} \rho_{12}],\\
\rho_{24}^{out} = Tr_{13}[U_{cl}\rho_{12}] 
\end{split}
\end{equation}
are separable.
\end{itemize}
We have considered the diagonal non-local output states to be $\rho_{14}^{out}$ and $\rho_{23}^{out}$ for the local case and the horizontal pairs $\rho_{12}^{out}$ and $\rho_{34}^{out}$ for the non-local case. It is important to mention that one could have chosen the vice-versa pairs as well for both cases.
\end{definition}
\begin{figure}[htbp]
\centering
        \begin{subfigure}[b]{0.22\textwidth} \includegraphics[scale=0.25]{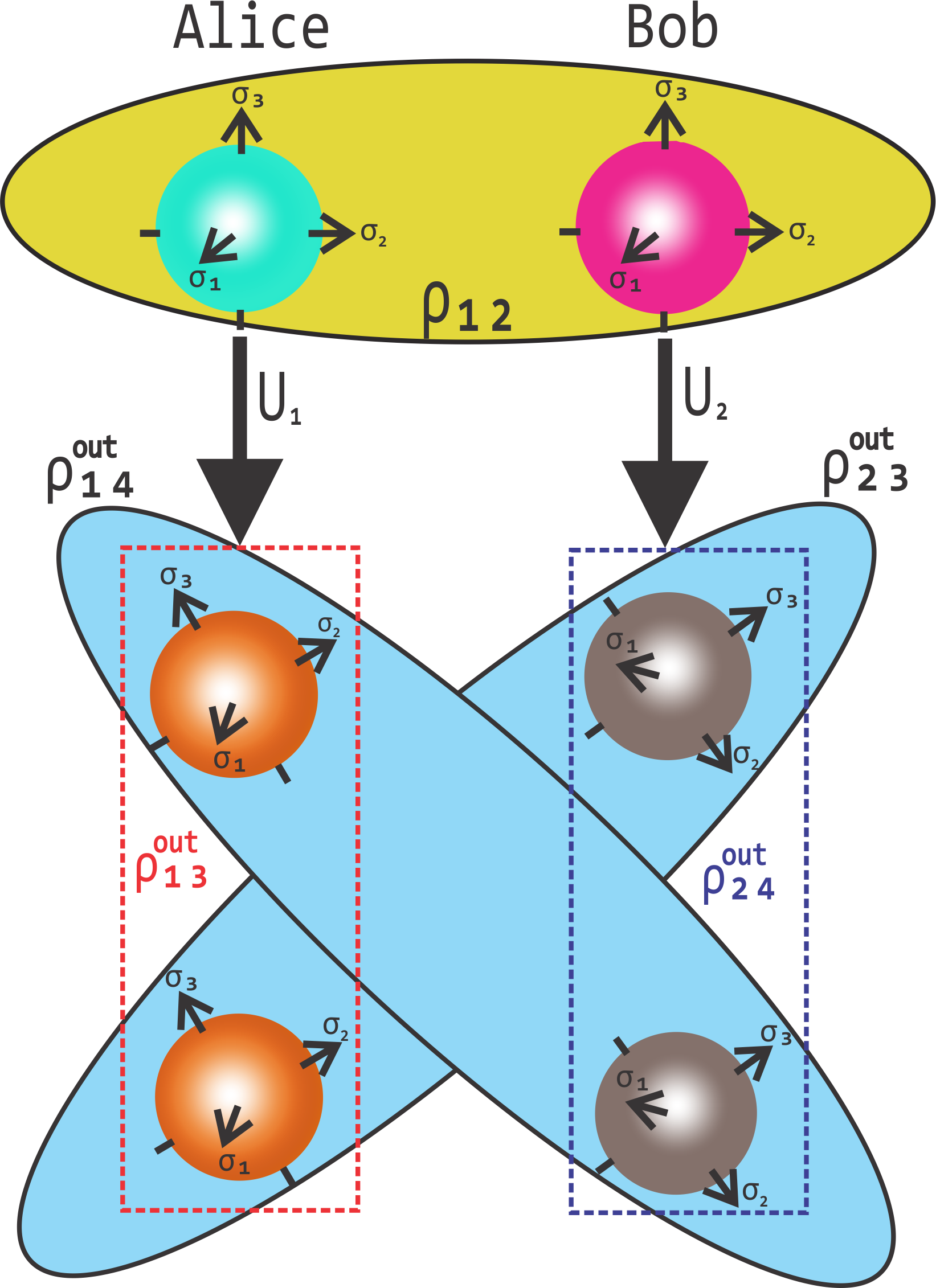}
                \caption{\scriptsize Local broadcasting. Only the diagonal nonlocal output pairs ($\rho^{out}_{14}$ and $\rho^{out}_{23}$) have been highlighted for clarity.}
                \label{fig:1to2loc}
        \end{subfigure}\hspace{0.01\textwidth}%
        \begin{subfigure}[b]{0.22\textwidth}
            \includegraphics[scale=0.25]{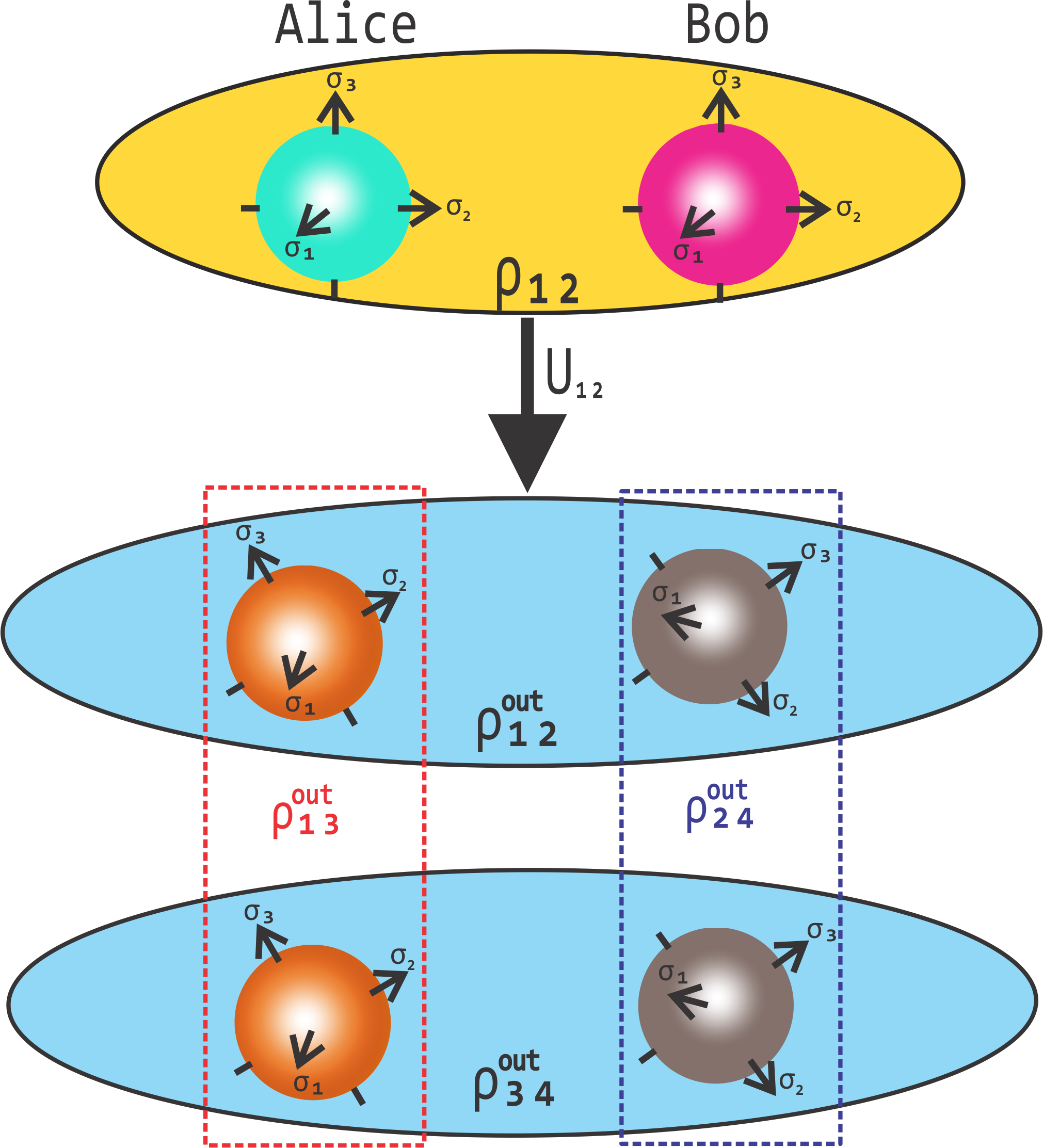}
                \caption{\scriptsize Nonlocal broadcasting. Only the horizontal nonlocal output pairs ($\rho^{out}_{12}$ and $\rho^{out}_{34}$) have been highlighted for clarity.}
                \label{fig:1to2nloc}
        \end{subfigure}%
        \caption{\scriptsize The figures depict broadcasting of entanglement. The boxes with dotted boundary highlight the local output pairs - $\rho^{out}_{13}$ and $\rho^{out}_{24}$. The difference in color and orientation of the qubits have been introduced to denote that the local pairs will be non-identical. However, when the input is a non-maximally entangled state, the outputs on Alice's side are same as that on Bob's side}\label{fig:localNonLocalCloning}
\end{figure}
The necessary and sufficient condition for entanglement detection in $2\otimes2$ and $2\otimes3$ systems is given by Peres-Horodecki criteria \cite{peres1996separability, horodecki1996m}. It states that for a state to be inseparable, at least one of the eigenvalue of the partially transposed state $\rho^T_{m\mu,n\nu} = \rho_{m\nu,n\mu}$ should be negative. If all the eigenvalues of the partially transposed state are positive, then the state is separable. This can be equivalently expressed by the condition that the value of at least one of the two determinants,

\[
W_{4} = 
  \begin{bmatrix}
    \rho_{00,00} & \rho_{01,00} & \rho_{00,10} & \rho_{01,10}\\
    \rho_{00,01} & \rho_{01,01} & \rho_{00,11} & \rho_{01,11}\\
\rho_{10,00} & \rho_{11,00} & \rho_{10,10} & \rho_{11,10}\\
\rho_{10,10} & \rho_{11,01} & \rho_{10,11} & \rho_{11,11}\\
  \end{bmatrix}
\] and
\[
W_{3} = 
  \begin{bmatrix}
    \rho_{00,00} & \rho_{01,00} & \rho_{00,10} \\
    \rho_{00,01} & \rho_{01,01} & \rho_{00,11} \\
\rho_{10,00} & \rho_{11,00} & \rho_{10,10}\\

  \end{bmatrix}
\] is negative, with
\[
W_{2} = 
  \begin{bmatrix}
    \rho_{00,00} & \rho_{01,00}  \\
    \rho_{00,01} & \rho_{01,01}  \\
  \end{bmatrix}
\] being simultaneously non-negative.



\subsubsection{Bell-diagonal state}
We consider the broadcasting of Bell-diagonal class, which is given by $\rho_{12}^{b} = p_1\ketbra{\Psi_+}{\Psi_+} + p_2\ketbra{\Psi_-}{\Psi_-} + p_3\ketbra{\phi_+}{\phi_+} + p_4\ketbra{\phi_-}{\phi_-} $, where $p_1, p_2, p_3$ and $p_4$ are classical mixing parameters \cite{bell-diagonal2010}. Here, $\ket{\Psi_{\pm}}, \ket{\phi_{\pm}}$ are Bell states. In terms of Bloch vectors and correlation matrix,  Bell-diagonal states can be expressed as,
\begin{equation}
\rho_{12}^{b} = \big\{\vec{0},\vec{0},T^b\big\},
\end{equation}
where $\vec{0}$ is the null matrix and $T^b$ = $\text{diag}[c_1, c_2, c_3]$ with ($-1\leq c_i\leq1$) is the correlation matrix.

We want to broadcast the entanglement in bell-diagonal state by the means of cloning. We can use seven types of cloning transformations, namely, orthogonal state independent local (OSIL), non-orthogonal state independent local (NOSIL), orthogonal state dependent local (OSDL), non-orthogonal state dependent local (NOSDL), orthogonal state independent non-local (OSINL),  orthogonal state dependent non-local (OSDNL) and non-orthogonal state dependent non-local (NOSDNL). This forms a nearly exhaustive list of cloning transformations. All the seven transformations are variants of orthogonal or non-orthogonal cloning procedures. 

\begin{figure}[h]
\begin{center}
\includegraphics[height=5.5cm,width=5.2cm]{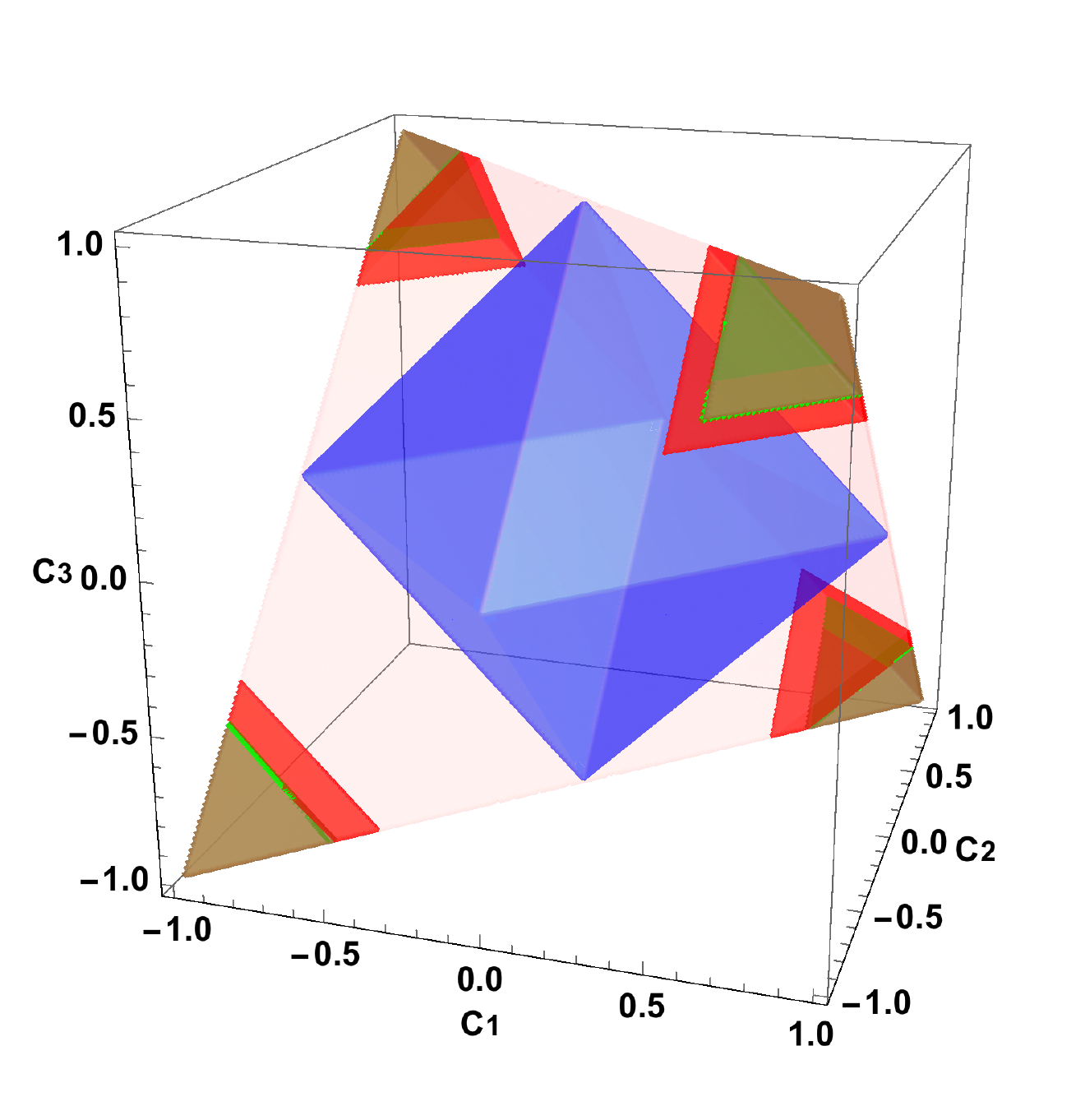} \includegraphics[height=2.5cm,width = 2cm]{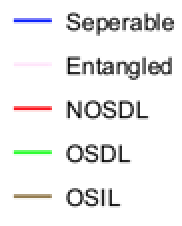}
\end{center}
\caption{\noindent \scriptsize
Plot comparing the performance of various local cloners in broadcasting of entanglement from a Bell-diagonal state. We highlight with a brown hue for denoting the zone when OSIL cloner was used, red hue for the use of NOSDL cloner and green hue for the case when OSDL cloner was employed.}
\label{fig:bellLocal}
\end{figure}

Let us approach the broadcasting of Bell-diagonal states in two parts: with local and non-local cloning. In the first part, we restrict ourselves to use of only local cloning transformations. When we talk about broadcasting using local cloners, we want the local output states, (as shown in Fig.~\ref{fig:localNonLocalCloning}), $\rho_{13}$ and $\rho_{24}$ to be separable and non-local outputs $\rho_{14}$ and $\rho_{23}$ to be entangled. 
In Fig.~\ref{fig:bellLocal}, we have shown the performance of various local cloning transformations for the task of broadcasting entanglement from the Bell-diagonal class of states. We have input state parameters $c_1$, $c_2$, $c_3$ on $x$, $y$ and $z$ axes, respectively. The pink colored tetrahedron represents the geometry of the Bell-diagonal state space; while the blue colored octahedron in the centre represents all separable states. Since entanglement cannot be generated using LOCC operations \cite{locc}, the set of Bell-diagonal states which can be broadcast using local cloning would only be a subset of the pink region outside the blue octahedron. There are three regions marked by brown, green and red colors. Since the state independent version of the orthogonal and the non-orthogonal cloners perform the same, they are represented together by the brown coloured region. The red colored region on the corners represents the performance by NOSDL cloner in broadcasting of entanglement, where the machine parameter $\mu = 0.7$ (which is the maximum value allowed by the Schwartz inequality). The red region is a super set of the brown region, and thus can be seen to exceed beyond the brown region. The green region is also a super set of the brown zone. It shows the performance of OSDL cloner, where machine parameter ($d = \sqrt{\frac{10}{55}}$). Clearly the best performing cloner is the NOSDL (i.e. the red one).
\begin{figure}[h]
\begin{center}
\includegraphics[height=5.5cm,width=5.2cm]{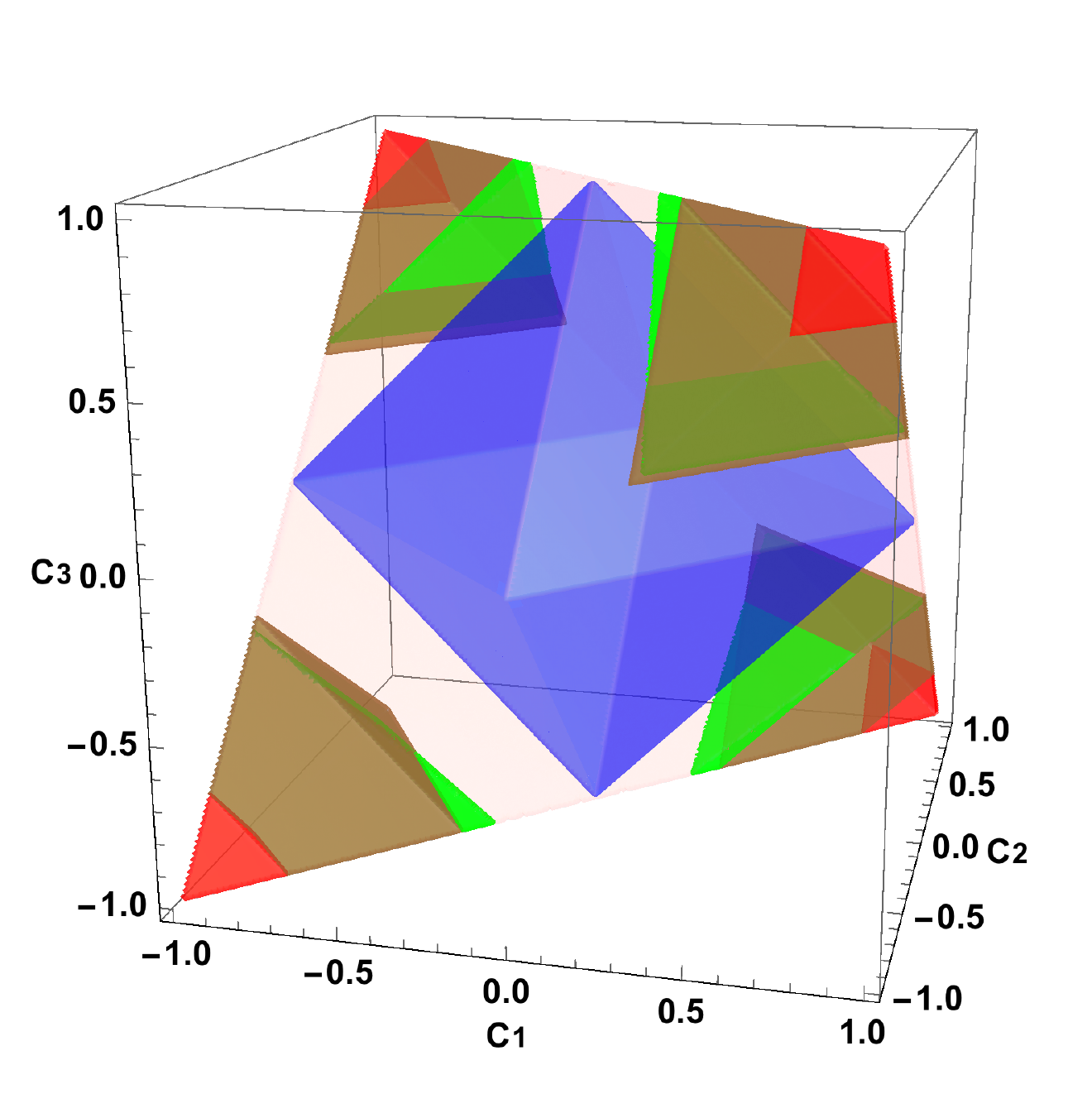}
\includegraphics[height=2.5cm,width = 2cm]{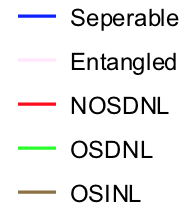}
\\
\end{center}
\caption{\noindent \scriptsize
Plot comparing the performance of various non-local cloners in broadcasting of entanglement from a Bell diagonal state. We highlight with a brown hue for denoting the zone when OSINL cloner was used, red hue for the use of NOSDNL cloner and green hue for the case when OSDNL cloner was employed.}
\label{fig:bellNonLocal}
\end{figure}

Now in the second part, for non-local cloning approaches, the non-local outputs are given by Eqs.(\ref{eq:nonlocalorthooutput}, \ref{eq:nonlocalnonorthooutput}) for the orthogonal and the non-orthogonal cases, respectively. 

The analysis of performance by various non-local cloners for broadcasting is shown in Fig.~\ref{fig:bellNonLocal}. We use the same labeling convention as the previous Fig.~\ref{fig:bellLocal}; however because there is no non-orthogonal universal non-local cloner, the brown region here only depicts the performance of OSINL cloner. The machine parameters here become $\mu = 0.40$ and $d = \sqrt{\frac{1}{15}}$.
It is interesting to note that the NOSDNL cloner performs the worst. This is because of the stricter Schwartz inequality constraints. The brown and green regions both contain some states which are not included in the other. So, there is no outright winner. However, one can choose the best cloner among the two, when they have prior information about the state. 
 
\section{Conclusion}
In a nutshell, in this work, we re-conceptualize the notion of state dependence in quantum cloning. Based on that, we introduce new state dependent cloning machines, which outperform their state independent counterparts, when prior partial information about the system to be cloned is available. Further, we have defined orthogonal cloners which can be prepared by ensuring that the machine states remain orthogonal, and non-orthogonal cloners where the machine states need not necessarily be orthogonal.
Such constructions were shown to be better than state independent cloning in the context of broadcasting of entanglement by taking the example of  Bell-diagonal class of states (BDS). These new types of state dependent cloning machines open up new possibilities for generating more number of lesser entangled pairs from a given entangled state. As an outlook, in this light it would be interesting to study broadcasting using asymmetric state dependent cloners \cite{ghiu2003asymmetric}, both locally and non locally. Also, it would be of interest to study broadcasting of coherence and discord using these type of state dependent cloners.

\section*{\label{sec:level1}ACKNOWLEDGEMENT}
S.C. thanks Prof. Mark M. Wilde for his insightful suggestions. S.C. acknowledges the internship grant from Erlangen Graduate School in Advanced Optical Technologies (SAOT) for supporting the research work as an intern at IIIT, Hyderabad \& HRI, Allahabad  in India.

\end{document}